\documentclass[a4paper,cits]{PoS}
\usepackage{amsmath,amssymb,amsfonts,amsthm,latexsym} 

\usepackage[normalem]{ulem}
\usepackage[utf8]{inputenc}
\usepackage[T1]{fontenc}

\usepackage{cancel}
\usepackage{graphicx}
\usepackage{latexsym}
\usepackage{bbm}
\usepackage{epsfig}
\usepackage{epstopdf}
\usepackage{subcaption}
\usepackage{color}
\usepackage{comment}

\usepackage{slashed}
\usepackage{placeins}
\usepackage{cleveref}
\usepackage{setspace}
\usepackage{url}

\usepackage{lmodern}
\usepackage{exscale}
\usepackage{tabularx}
\usepackage{xcolor}
\usepackage{tikz,wrapfig}
\usepackage{enumitem}
\usepackage{mcite}
\usetikzlibrary{fadings,calc,decorations.markings}

\makeatletter
\newsavebox{\@brx}
\newcommand{\llangle}[1][]{\savebox{\@brx}{\(\m@th{#1\langle}\)}%
  \mathopen{\copy\@brx\kern-0.5\wd\@brx\usebox{\@brx}}}
\newcommand{\rrangle}[1][]{\savebox{\@brx}{\(\m@th{#1\rangle}\)}%
  \mathclose{\copy\@brx\kern-0.5\wd\@brx\usebox{\@brx}}}
\makeatother

\makeatletter
\def\namedlabel#1#2{\begingroup
    #2%
    \def\@currentlabel{#2}%
    \phantomsection\label{#1}\endgroup
}
\makeatother

\newcommand{\dd}{\textrm{d}}
\newcommand{\Tr}{\textrm{Tr}\,}
\newcommand{\AS}{\alpha_\mathrm{s}}

\title{Static force from the lattice}

\ShortTitle{Static force from the lattice}

\author{Nora Brambilla$^{a\ddagger}$, \speaker{Viljami Leino}$^{a\ddagger}$, Owe Philipsen$^b$, Christian Reisinger$^b$, Antonio~Vairo$^{a\ddagger}$, and Marc Wagner$^b$ \\
    $^a$ Technische Universit\"at M\"unchen, Physik Department, James-Franck-Str.\ 1, 85748 Garching, Germany \\
    $^b$ Goethe Universit\"at Frankfurt am Main, Institut für Theoretische Physik, Max-von-Laue-Str.\ 1, 60438 Frankfurt am Main, Germany \\
 		E-mail: \email{nora.brambilla@ph.tum.de}, \email{viljami.leino@tum.de}, \email{philipsen@itp.uni-frankfurt.de},
		\email{reisinger@itp.uni-frankfurt.de}, \email{antonio.vairo@ph.tum.de}, \email{mwagner@itp.uni-frankfurt.de}}
\author{$^\ddagger$(TUMQCD collaboration)}

\abstract{We present a novel approach to compute the force between a static quark and a static antiquark from lattice gauge theory directly, 
          rather than extracting it from the static energy. 
		  We explore this approach for SU(3) pure gauge theory using the multilevel algorithm and smeared operators.
		 }

\FullConference{37th International Symposium on Lattice Field Theory - Lattice2019\\
		16-22 June 2019\\
		Wuhan, China}

\begin{document}


\section{Introduction}

The strong coupling $\AS$ can be measured quite accurately by matching observables computed in perturbation theory and lattice QCD~\cite{Pich:2018lmu,Aoki:2019cca}. 
Among different methods, the extraction of $\AS$ from the energy, 
$E(r)$ of a static quark-antiquark pair located at a distance $r$, 
can be done both on the lattice and perturbatively.

The static energy $E(r)$, which is an important quantity in pure gauge theory as well as in QCD, 
can be determined using lattice gauge theory by computing rectangular Wilson loops with spatial extent $r$ and large temporal extent $T$,
\begin{align}\label{EQN_V} 
E(r) = -\lim_{T \rightarrow \infty} \frac{\ln \langle \textrm{Tr}(W_{r \times T}) \rangle}{T} \,,
\quad \quad W_{r \times T} = P\bigg\{\exp\bigg(i \oint_{r \times T} dz_\mu \, gA_\mu\bigg)\bigg\} \,,
\end{align}
where $P\{ \ldots \}$ stands for path ordering.
On the other hand the perturbative expression for $E(r)$, 
which is valid for small separations $r \ll 1 / \Lambda_\mathrm{QCD}$ and $\AS \ll 1$, is known up to N$^3$LL order~\cite{Brambilla:2009bi},
\begin{align}
E(r) = \Lambda - C_\mathrm{F}\frac{\AS}{r}
\Bigg\{1&+\frac{\AS}{4\pi}\tilde{a}_1+\left(\frac{\AS}{4\pi}\right)^2\tilde{a}_2 
+\left(\frac{\AS}{4\pi}\right)^3\left[a_3^\mathrm{L}\log\frac{C_\mathrm{A}\AS}{2}+\tilde{a}_3\right] 
\nonumber \\&+
\left(\frac{\AS}{4\pi}\right)^4
\left[a_4^\mathrm{L2}\log^2\frac{C_\mathrm{A}\AS}{2}+a_4^\mathrm{L}\log\frac{C_\mathrm{A}\AS}{2}+\tilde{a}_4\right]
\Bigg\} \, ,
\end{align}
where $\Lambda$ is a constant, $\AS\equiv \AS(1/r)$, $C_\mathrm{F}=4/3$ and $C_\mathrm{A}=3$ for SU(3), and, 
the constants $\tilde{a}_k,\,a^\mathrm{L}_k$ are summarized in Ref.~\cite{Tormo:2013tha}. 
See also Ref.~\cite{Tormo:2013tha} for references to the original literature.
The lattice result and the perturbative result for $E(r)$ differ by a constant related to the self-energy. 
In lattice gauge theory the self-energy diverges as $1/a$ ($a$ denotes the lattice spacing), 
while in perturbation theory, when using dimensional regularization, there is a renormalon ambiguity.

The different but constant self energies can be eliminated by taking the spatial derivative of the static energy, 
i.e.,~by considering the static force $F(r) = \partial_r E(r)$. This approach has a long history~\cite{Booth:1992bm}, 
but taking numerical derivatives of the lattice result for the static energy typically increases numerical errors. 
In this work we, thus, explore a different approach, which was proposed in Ref.~\cite{Vairo:2015vgb,*Vairo:2016pxb}, 
the direct lattice computation of the static force using
\begin{align}\label{EQN_F} 
F(r) = -\lim_{T \rightarrow \infty} \frac{i}{\langle \textrm{Tr}(W_{r \times T}) \rangle} \bigg\langle \textrm{Tr}\bigg(P\bigg\{\exp\bigg(i \oint_{r \times T} \dd z_\mu 
\, gA_\mu\bigg) \hat{\mathbf{r}} \cdot g \mathbf{E}(\mathbf{r},t^\ast)\bigg)\bigg\} \bigg\rangle\,,
\end{align}
where $\hat{\mathbf{r}}$ is the direction of the separation of the static color charges and $\mathbf{E}(\mathbf{r},t)$ 
denotes the (Euclidean) chromoelectric field located on one of the temporal Wilson lines at $-T/2 < t < T/2$. 
The right hand side of~\eqref{EQN_F} is independent of $t^\ast$ as long as $t^\ast$ is 
a fixed time. 

\section{Derivation of Eq.\ (\ref{EQN_F})}

Equation~\eqref{EQN_F} was first derived in~\cite{Brambilla:2000gk}. 
In the following, we illustrate the derivation 
in the Abelian case, where traces and path ordering can be ignored. 
The generalization to the non-Abelian case is straightforward and can be found in detail in Ref.~\cite{Eichberg2019}.

We start by applying the spatial derivative to the static energy, which can be expressed in terms of Wilson loops as done in Eq.\ (\ref{EQN_V}),
\begin{align}\label{EQN_001} 
F(r) = \partial_r E(r) = -\lim_{T \rightarrow \infty} \frac{\langle \partial_r W_{r \times T} \rangle}{T \langle W_{r \times T} \rangle} \, .
\end{align}
The spatial derivative of the Wilson loop can be written as a finite difference,
\begin{align}
\label{EQN_002}\partial_r W_{r \times T} = 
\lim_{\Delta \rightarrow 0} \frac{1}{\Delta} \Big(W_{(r+\Delta) \times T} - W_{r \times T}\Big) = 
W_{r \times T} \lim_{\Delta \rightarrow 0} \frac{W_{\Delta \times T} - 1}{\Delta} \, ,
\end{align}
and
\begin{align}
\nonumber
\lim_{\Delta \rightarrow 0} \frac{W_{\Delta \times T} - 1}{\Delta} &
= \lim_{\Delta \rightarrow 0} \frac{1}{\Delta} \bigg(\exp\bigg(i \oint_{\Delta \times T} \dd z_\mu \; gA_\mu\bigg) - 1\bigg) = \\
\nonumber &
= \lim_{\Delta \rightarrow 0} \frac{1}{\Delta} \bigg(\exp\bigg(\frac{i}{\Delta} \int_{-\frac{T}{2}}^{\frac{T}{2}} \dd t \; 
\oint_{\Delta \times \Delta} dz_\mu \; gA_\mu\bigg) - 1\bigg) = \\
\label{EQN_003} &
= \lim_{\Delta \rightarrow 0} \frac{1}{\Delta} \bigg(\exp\bigg(i \Delta \int_{-\frac{T}{2}}^{\frac{T}{2}} \dd t \; \hat{\mathbf{r}} \cdot g\mathbf{E}(\mathbf{r},t)\bigg) -1\bigg)
= i \int_{-\frac{T}{2}}^{\frac{T}{2}} \dd t \; \hat{\mathbf{r}} \cdot g \mathbf{E}(\mathbf{r},t) \, ,
\end{align}
where $\displaystyle \lim_{\Delta \rightarrow 0} (1 / \Delta^2) \oint_{\Delta \times \Delta} \dd z_\mu \; g A_\mu 
= \hat{\mathbf{r}} \cdot g\mathbf{E}(\mathbf{r},t)$ has been used. Combining Eqs.\ (\ref{EQN_001}), (\ref{EQN_002}) and (\ref{EQN_003}) leads to
\begin{align}
F(r)
= -\lim_{T \rightarrow \infty} \frac{i}{T \langle W_{r \times T} \rangle} \bigg\langle W_{r \times T} \int_{-\frac{T}{2}}^{\frac{T}{2}} \dd t \; 
\hat{\mathbf{r}} \cdot g\mathbf{E}(\mathbf{r},t) \bigg\rangle
= -\lim_{T \rightarrow \infty} \frac{i}{\langle W_{r \times T} \rangle} \Big\langle W_{r \times T} \, \hat{\mathbf{r}} \cdot g\mathbf{E}(\mathbf{r},t^\ast) \Big\rangle \, ,
\end{align}
which is the Abelian version of Eq.~\eqref{EQN_F}.
The last equality holds when the limit $T\rightarrow \infty$ is taken for a specific choice of $t^\ast$. 
In actual simulations with finite $T$ we use the expression on the right hand side of this equality, 
because corrections to the $T\rightarrow \infty$ result are exponentially suppressed with respect to $T$, 
while for the expression on the left hand side corrections may
be only $1/T$ suppressed.

\section{Details of the lattice computation}

In our exploratory study, we compute the static force in pure SU(3) lattice gauge theory 
by evaluating the right hand side of Eq.~\eqref{EQN_F} on the lattice.
This is done by inserting a chromoelectric field to a Wilson loop in a gauge invariant way~\cite{Bali:1997am,Koma:2006fw}.
We perform this calculation in two different ways:
\begin{enumerate}[label={\textbf{(\arabic*)}}]
\item \label{item:1} We consider lattices with large temporal extent $T$ and represent the closed loops $\displaystyle \oint_{r \times T} \ldots$ 
by two Polyakov loops (i.e.,~omitting the spatial parallel transporters) using the Butterfly definition for the chromoelectric field: 
\begin{align}\label{eq:butterfly}
E_i = \frac{1}{2} \Big(F_{0i} + F_{-i0}\Big)\,, \qquad F_{\mu\nu} = \frac{1}{2 i g a^2} \Big(P_{\mu,\nu} - P^\dagger_{\mu,\nu}\Big) \, ,
\end{align}
where $P_{\mu,\nu}(x) = U_\mu(x) U_\nu(x+\hat{\mu}) U^\dagger_\mu(x+\hat{\nu}) U_\nu^\dagger(x)$ denotes the plaquette. 
The expectation values are computed using the multilevel algorithm~\cite{Luscher:2001up} with $4$ sublattices and $6\,000$ sub-updates, 
where we have generated around $100$ gauge link configurations for each ensemble. 

To reduce discretization errors,
we redefine the separation $r$ such that the tree-level results of $F(r)$ from lattice and continuum perturbation theory match~\cite{Necco:2001xg}, 
i.e.~$r \rightarrow r_I(r)$, where $r_I(r)$ is defined via
\begin{align}
\frac{1}{4 \pi r_I^2} = \frac{G(r+a) - G(r-a)}{2 a}\,,
\end{align}
with the lattice gluon propagator in coordinate space
\begin{align}
G(r) = \frac{1}{a} \int_{-\pi}^{+\pi} \frac{\dd^3k}{(2\pi)^3} \frac{\prod_{j=1}^{3} \cos(x_j k_j/a)}{4 \sum_{j=1}^{3} \sin^2(k_j/2)} \, .
\end{align}

\item \label{item:2} We evaluate 
Eq.~\eqref{EQN_F} 
by choosing $t^\ast$ as close to zero as possible and using the Clover definition for the chromoelectric field: 
\begin{align}\label{eq:clover}
E_i = \frac{1}{2 i g a^2} \Big(\Pi_{i0} - \Pi_{i0}^\dagger\Big)\,, \quad \quad \Pi_{\mu\nu} = \frac{1}{4} \Big(P_{\mu,\nu} + P_{\nu,-\mu} + P_{-\mu,-\nu} + P_{-\nu,\mu}\Big) \, .
\end{align}
We employ 50 APE-smearing steps on the links forming the spatial parallel transporters of the closed loops 
$\displaystyle \oint_{r \times T} \ldots$ to maximize the ground state overlap. 
The expectation values are computed on $400$ gauge link configurations separated by 30 updates each. 
The $T \rightarrow \infty$ limit is approximated by 
temporal separations $T \geq 8 \, a$.
\end{enumerate}


\section{Renormalization}\label{sec:HM}

The chromoelectric field appearing in Eq.\ \eqref{EQN_F} requires renormalization. 
For this we use the Huntley and Michael (HM) procedure~\cite{Huntley:1986de}, 
which removes self-energy contributions up to order $\mathcal{O}(g^4)$. 
In detail we multiply the chromoelectric field by 
$Z_\mathrm{E} = \Tr(W_{r \times T}) / \Tr(W_{r \times T}\,\bar{E})$,
where $\bar{\mathbf{E}}$ is given by:
\begin{alignat}{2}
\bar{E}_i &= \frac{1}{2} \Big(\bar{F}_{0i} + \bar{F}_{-i0}\Big) \,, &\qquad \bar{F}_{\mu\nu} &= \frac{1}{2 i g a^2} \Big(P_{\mu,\nu} + P^\dagger_{\mu,\nu}\Big) \,,\\
\intertext{or}
\bar{E}_i &= \frac{1}{2 i g a^2} \Big(\Pi_{i0} + \Pi_{i0}^\dagger\Big) \,, &\qquad \Pi_{\mu\nu} &= \frac{1}{4} \Big(P_{\mu,\nu} + P_{\nu,-\mu} + P_{-\mu,-\nu} + P_{-\nu,\mu}\Big)
\,,
\end{alignat}
for our two computations \ref{item:1} and \ref{item:2}, respectively.


\section{Numerical results}

We have performed simulations for several values of the lattice spacing and several lattice volumes, 
where we relate the coupling constant $\beta$ and the lattice spacing $a$ via~\cite{Necco:2001xg}:
%
\begin{align}\label{eq:sc}
\ln(a/r_0) = -1.6804-1.7331 \, (\beta-6)+0.7849 \, (\beta-6)^2-0.4428 \, (\beta-6)^3 \, .
\end{align}

\begin{figure}[t]
  \includegraphics[width=0.49\textwidth]{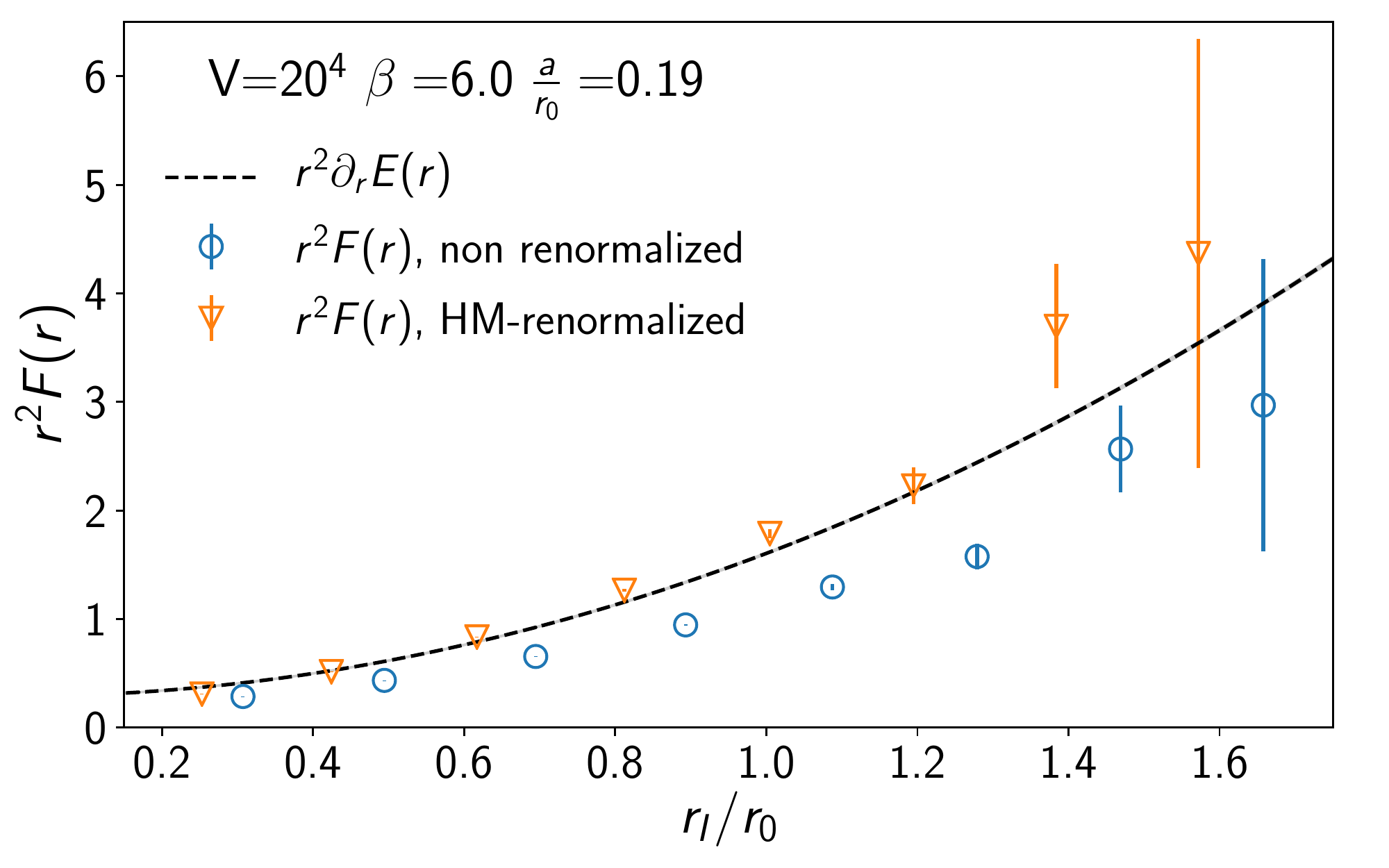}
  \includegraphics[width=0.49\textwidth]{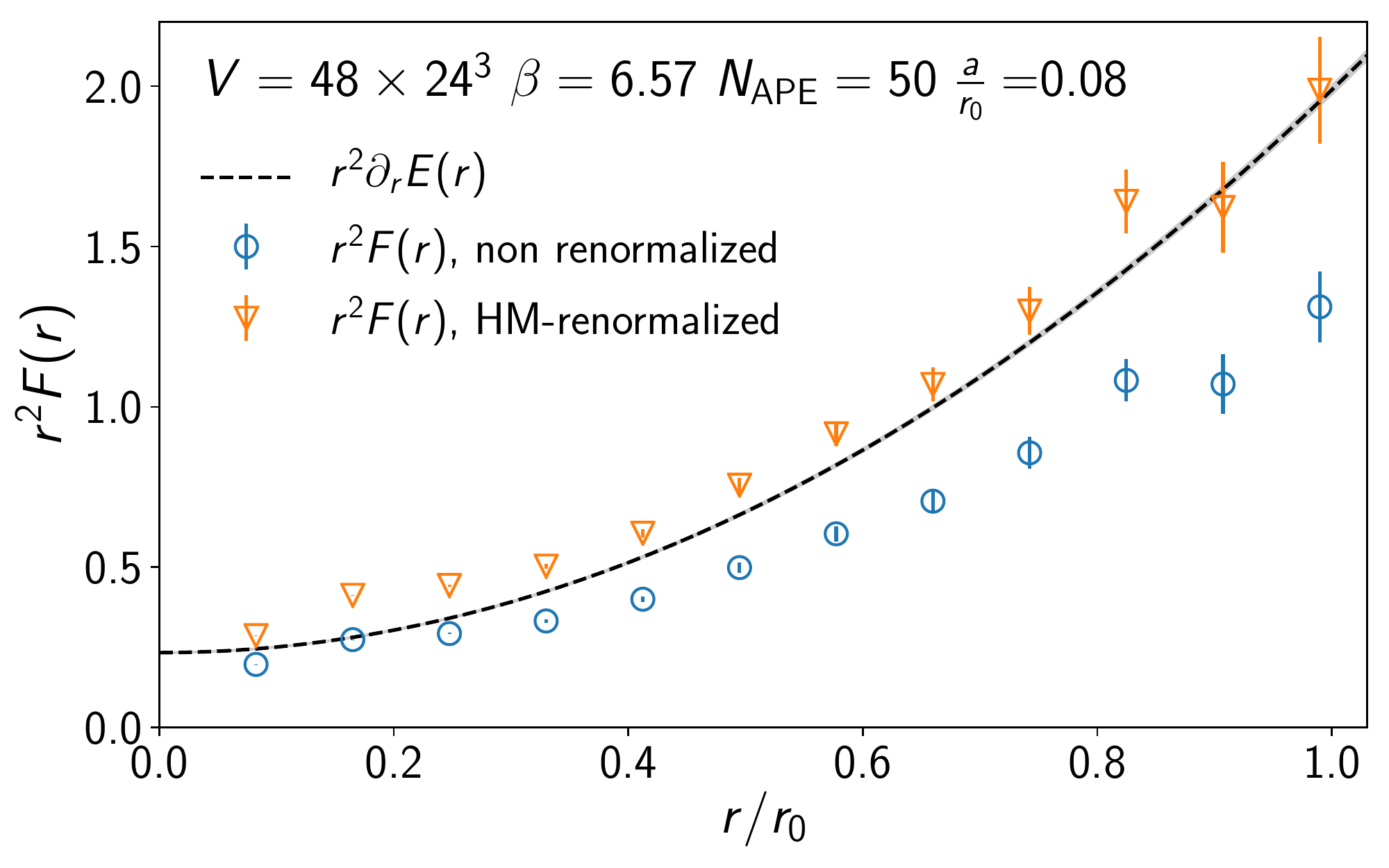}
  \caption[b]{Lattice results of the direct computation of the static force via Eq.~\eqref{EQN_F}. 
              Left plot: computation \ref{item:1}
			  with $\beta=6.0$ (corresponding to $a= 0.19 \, r_0$) and lattice volume $20^4$.
			  Right plot: computation \ref{item:2} with $\beta=6.57$ (corresponding to $a= 0.08 \, r_0$) 
			  and lattice volume $48 \times 24^3$.}
  \label{fig:measure}
\end{figure}

Selected lattice results of the direct computation of the static force via Eq.~\eqref{EQN_F} are shown in Figure~\ref{fig:measure}
(left plot computation \ref{item:1}, right plot computation \ref{item:2})
in comparison to the force obtained by taking the derivative of the static energy. 
For the latter we fit a Cornell ansatz $E(r) = V_0 - \kappa/r + \sigma r$ to the lattice results for the static energy, 
which leads to $F(r) = \kappa/r^2 + \sigma$ (black dashed curves). 
As expected, we observe that there is better agreement with the HM renormalized data points for $r^2F(r)$ (orange dots) 
than with the non-renormalized data points for $r^2F(r)$ (blue dots), 
especially in the left plot where tree-level improvement is used.

\begin{figure}
\includegraphics[width=0.9\textwidth]{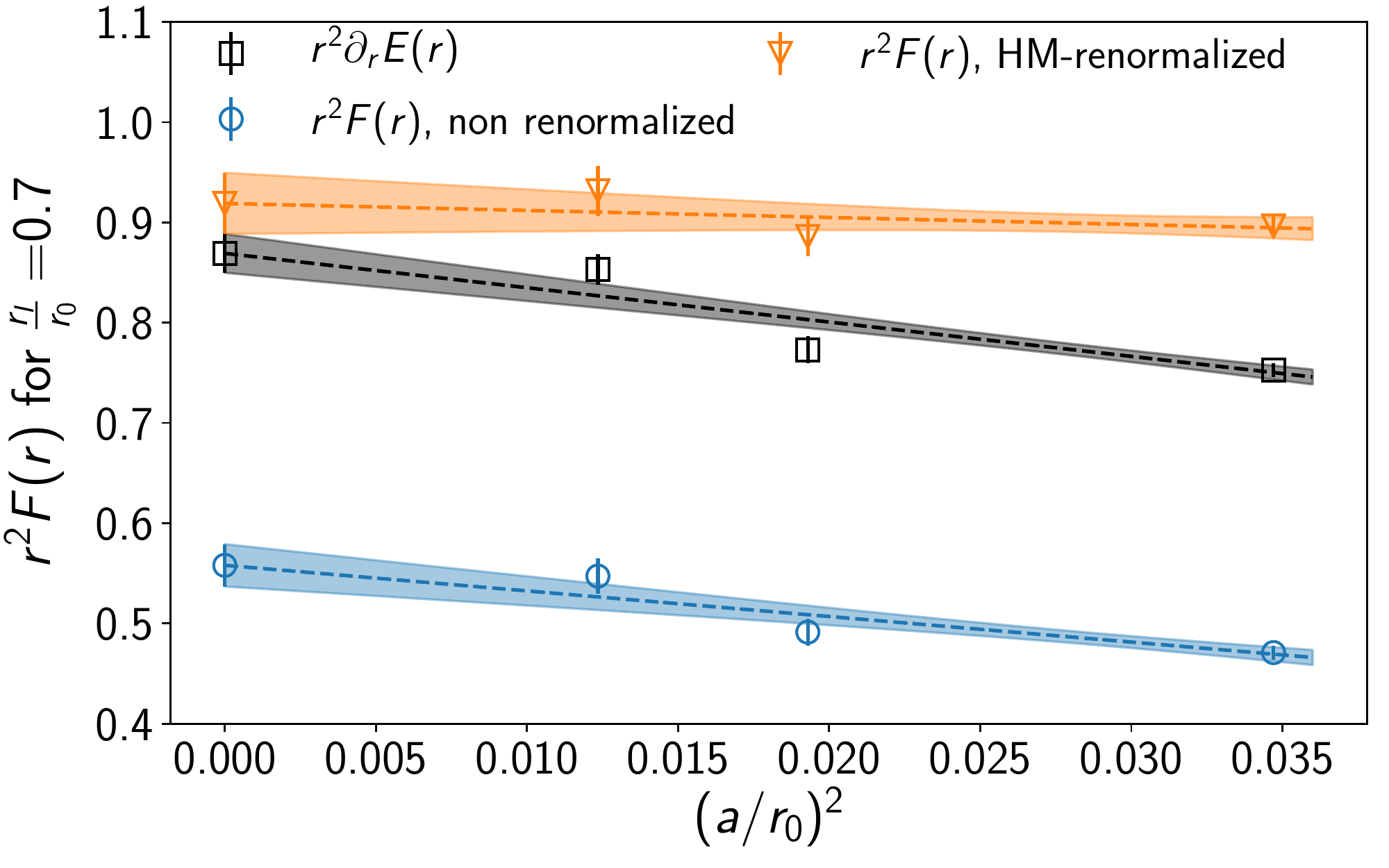}
\caption{Continuum limit of $r^2 F(r)$ for $r = 0.7 \, r_0$ 
         at constant physical volume of extent $2.28 \, r_0$ (computation \ref{item:1}).}
\label{fig:continuum}
\end{figure}

For computation \ref{item:1} we also study the continuum limit at constant physical volume of extent $2.28 \, r_0$ 
using lattice volumes $12^4$, $16^4$, and $20^4$ with 
$\beta = 6.0$, $6.183$, $6.3406$ obtained from Eq.~\eqref{eq:sc}. 
We interpolate the data with cubic splines and 
in Figure~\ref{fig:continuum} we show the continuum limit of $r^2 F(r)$ for a single value of the (improved) separation 
$r \equiv r_I \simeq 0.7 \, r_0$.
Even though our statistical precision is at the moment still limited, 
we observe clear indication that the HM renormalized result for $F(r)$ agrees 
with the derivative of $E(r)$ in the continuum limit, while the non-renormalized result for $F(r)$ is significantly different.
 
%
%


\FloatBarrier
\acknowledgments{
We acknowledge useful conversations with Michael Eichberg and Saumen Datta for sharing his multilevel code with us~\cite{Banerjee:2011ra}.
N.B.,~V.L.,~and~A.V.~acknowledge support from the DFG cluster of excellence ORIGINS
\href{www.origins-cluster.de}{(www.origins-cluster.de)}.
C.R.\ acknowledges support by a Karin and Carlo Giersch Scholarship of the Giersch foundation. 
M.W.\ acknowledges funding by the Heisenberg Programme of the Deutsche Forschungsgemeinschaft 
(DFG, German Research Foundation) -- Projektnummer 399217702.
The simulations have been carried out on the computing facilities of the Computational 
Center for Particle and Astrophysics (C2PAP) at the Technical University of Munich, 
and on the Goethe-HLR high-performance computer of the Frankfurt University. 
We would like to thank HPC-Hessen, funded by the State Ministry of Higher Education, Research and the Arts, for programming advice.
The simulations were performed with Chroma~\cite{Edwards:2004sx} and modified version of the multilevel code
developed for~\cite{Banerjee:2011ra}.
This research was supported by the Munich Institute for 
Astro- and Particle Physics (MIAPP) of the DFG Excellence Cluster ORIGINS.
This work was supported in part by the Helmholtz International Center for FAIR within the framework of the LOEWE program launched by the State of Hesse.
}

\bibliographystyle{jhep_modified}
\bibliography{proceedings.bib}

%

\end{document}